\documentstyle[12pt]{article}
 
\def\pa{\partial}

 \def\G{\Gamma}

\def\d{\delta}

\def\l{\lambda} \def\L{\Lambda}
\def\m{\mu}

\def\x{\chi}
\def\p{\phi} \def\P{\Phi}

\def\loon{\relbar\joinrel\relbar\joinrel\relbar\joinrel\rightarrow}
\def\loonl{\relbar\joinrel\relbar\joinrel\relbar\joinrel\relbar
\joinrel\rightarrow}
\def\ul{\underline}
\def\ni{\noindent}

\renewcommand{\baselinestretch}{1}

\begin{document}

\begin{flushright}
hep--ph/9311319\\
BRX--TH--351 -- rev.
\end{flushright}

\vspace{.5in}

\begin{center}
{\Large{\bf THE CUTOFF $\l\p^4$ O(N) MODEL IN
\\ \vspace{.1in} THE LARGE N LIMIT}}

\normalsize

\vspace{.3in}

Jo\~ao P. Nunes\footnote{Supported by JNICT--PROGRAMA
CI\^{E}NCIA -- BD/1414/91 -- RM.}\\

\vspace{.1in}

and\\

\vspace{.1in}

Howard J. Schnitzer\footnote{Supported in part by the DOE under
grant
DE--FG02--92ER40706.}\\

\vspace{.1in}

Martin Fisher School of Physics\\
Brandeis University\\
Waltham, MA 02254

\vspace{.2in}

{\bf Abstract}
\end{center}

A cutoff version of the $\l\p^4$ O(N) model is considered to leading
order in 1/N with particular attention to the effective potential, which
is surprisingly rich in structure.  With suitable restriction on a
background classical field, one finds a phenomenologically viable model
with spontaneously broken symmetry, a potential bounded below, and
amplitudes free of tachyons.  The model has an O(N--1) singlet
resonance in both weak and strong coupling, which can be interpreted
as the Higgs meson in applications.  Further, an unphysical resonance,
which can be used to define a triviality scale for the model,
appears at a mass above the cutoff mass $\L$.  The phenomelogical
aspects of our discussion are consistent with previous studies of closely
related models by Heller, {\it et al.}

The question of the double-scaling limit for the cutoff model is
considered as an application of the effective potential.  It is shown that
the double-scaling limit is not possible.

\vspace{1.3in}

\noindent November 1993

\newpage

\ni{\bf I. ~Introduction}

The possibility of a strongly interacting Higgs sector, with large Higgs
mass, suggests the need for methods which can deal with the
interactions of scalars that are non-perturbative in the coupling
constant.  One approach to this issue is to reorganize the theory in
terms of some other expansion parameter, such as the 1/N expansion
for a theory with internal symmetry such as O(4), continued to
O(N) for example.  Predictions are then obtained by evaluating results
of the expansion
at the physical value of N (N=4 say).  The 1/N expansion for
$\l \p^4$ theory (in 3+1 dimensions) with O(N) symmetry (the
so-called vector
model) has been extensively studied as a renormalized field theory
\cite{001,002,003,004}.
However, the renormalized vector model encounters a number of
problems \cite{001,002}.  Among these are:

\noindent 1)~ The effective potential of the theory is double valued
\cite{002},
with the lower
energy branch of the potential describing a phase of the theory with
\ul{unbroken} internal symmetry, {\it i.e.,} $<\P_a > = 0$.  This phase
is tachyon free in all orders of the 1/N expansion.  The higher energy
branch of the effective potential does allow a spontaneously broken
symmetry.  However, this phase contains tachyons, presumably as a
symptom of decay to the lower energy phase.  In higher
orders of the 1/N expansion, the higher
energy branch of the effective potential becomes everywhere complex.

\noindent 2)~ The effective potential has no lowest energy bound as the
external
field $\p \rightarrow \infty$ \cite{001,002,003}.  The tachyon-free
phase ({\it i.e.,} with
$< \P_a > =0$) of the 1/N expansion tunnels non-perturbatively to this
unstable vacuum, with an amplitude proportional to exp(--N).\\
3)~ Most importantly, it is widely believed \cite{003,005} that $\l\p^4$
theory in
4--dimensions is actually a trivial, free-field theory, which may be at the
root of the problems summarized above.\\
These difficulties would seem to make the renormalized vector model,
evaluated in the 1/N expansion, unsuitable for phenomenology.

One possible way to deal with these problems is to consider a
cutoff version of the vector model in the 1/N expansion \cite{006}.  One
introduces a cutoff $\L$ into the theory, which represents a
mass-scale above which the self-coupled Higgs sector can no longer be
considered isolated from other essential degrees of freedom of a more
complete theory.  There are other possible interpretations of the cutoff,
particularly if the Higgs is not elementary, but only represents a scalar
bound-state of some effective field theory.  However, one does not need
to commit
oneself to any particular physical interpretation of the cutoff in this
paper.

A number of results for the cutoff vector model in the large N limit are
available \cite{006,007,008}, but no systematic study of the effective
potential of the model
has been undertaken.
In this paper we give a careful presentation of the effective potential to
leading order in 1/N for the cutoff vector model.  This is the main new
contribution of this paper.  We find that the
effective potential is surprisingly rich in structure, with several phases
possible, depending on the parameters of the model.  Some restrictions
on the parameters and external field strengths will be required to
obtain a phenomenologically viable model for energies $<\L$.  The
nature of these restrictions will become clear after our analysis of the
effective potential.
With such restrictions we find a phenomenologically viable model with
spontaneously broken symmetry, a potential bounded below, and
amplitudes free of tachyons.
We then study the resonance structure of meson-meson scattering in
the \linebreak O(N--1) singlet sector in this
phase.  The scattering amplitude  exhibits a resonance which can be
interpreted as the  Higgs meson of the model, both in weak and in
strong coupling, as well as an unphysical resonance above the cutoff
mass $\L$, which we interpret as defining the triviality mass-scale of
the theory.

In Sec. 2 we formulate the cutoff version of the vector model, while Sec.
3 presents the effective potential of the model.  An analysis of the
Green's functions of the model in Sec. 4 provides criteria for the
presence or absence of tachyons.  Sec. 5 considers the O(N) scalar
bound-state structure in the phase with spontaneously broken
symmetry.
This allows us to study the dependence of a Higgs meson mass on the
coupling
constant and cutoff of the model.
In Sec. 6 we consider the possibility of a double-scaling limit for this
cutoff model.  Some important observations about cutoff models are
included in the concluding Sec. 7.

\vspace{.2in}

\ni{\bf II. ~The Cutoff Model}

We formulate the cutoff vector model using the conventions and
notation of Abbott, {\it et al.,} ref. \cite{002}.  The Lagrangian density
is
\renewcommand{\theequation}{2.\arabic{equation}}
\setcounter{equation}{0}
\begin{equation}
{\cal L} = \frac{1}{2} (\pa_\m \P )^2 -
\frac{1}{2} \; \m^2_0 \P^2 -
\frac{\l_0}{4!N} \; (\P^2)^2
\end{equation}
where $\P_a$ ($a$ = 1 to N) is an N-component quantum field, and
$\P^2 = \sum^N_{a=1} \; \P_a \P_a$.  In (2.1) $\m_0$ and $\l_0$ are
the bare mass and coupling constant, respectively.  To leading order in
1/N the effective potential satisfies
\begin{equation}
\frac{\pa V(\p^2)}{\pa\p_a} = \x \, \p_a
\end{equation}
with $\x$ related to the (constant) classical field $\p_a$ by the gap
equation
\begin{equation}
\x = \m^2_0 + \frac{1}{6} \, \l_0 \left( \frac{\p^2}{N} \right)
+ \frac{1}{6} \, \l_0 \, \int
\frac{d^4k}{(2\pi )^4} \: \frac{1}{k^2 +\x}
\end{equation}
where the integral is over Euclidean momenta.  In order to make the
gap equation well-defined, the integration in (2.3) is cutoff at $\L$.
With fixed cutoff, both $\m_0$ and $\l_0$ are finite.  Throughout this
paper we regard $\m_0, \; \l_0$ and $\L$ as fixed, finite parameters
of the theory.  In particular, we will \ul{not} renormalize the theory,
and will \ul{not} ``run" the cutoff $\L$ in the sense of the
renormalization group.  For us $\L$ is a fixed energy above which the
\ul{pure} $\l\p^4$ no longer should be considered as an isolated
theory.  As a consequence the gap
equation becomes
\begin{equation}
\x = \m^2_0 + \frac{\l_0}{6} \left( \frac{\p^2}{N} \right) +
\left( \frac{\l_0}{96\pi^2} \right) \;
\left[ \L^2 + \x \log \, \left( \frac{\x}{\L^2 +\x} \right) \right]
\end{equation}
It is important to note that we do \ul{not} assume $(\x /\L^2) \ll 1$,
contrary to what is frequently done.  One can use (2.2) and (2.4) to find
the effective potential
\begin{eqnarray}
V & = & - \: \frac{3}{2} \: \frac{N}{\l_0} \: \x^2 \; + \;
\frac{1}{2} \: \x \p^2 \; + \; \frac{3N\m^2_0}{\l_0} \: \x \nonumber
\\[.2in]
& & - \: \frac{N}{64\pi^2} \,
\left[ \L^4 \: \log \, \left( \frac{\L^2}{\L^2 +\x} \right) -
\x^2 \log \, \left( \frac{\x}{\L^2 +\x} \right)
- \x \, \L^2 \right]
\end{eqnarray}

It is easy to extract some general properties of the cutoff theory from
(2.2), (2.4) and (2.5).   Observe that for $\l_0 > 0$
\begin{equation}
\x   \;\;\; _{\stackrel{\loon}{ {\rm Re}\, \x \rightarrow +\infty}} \;\;\;
\frac{\l_0\p^2}{6N} \; + \; {\cal O}(1) \;\; ,
\end{equation}

\begin{equation}
{\rm Re} \;\; \frac{\pa V}{\pa \p^2} \;
_{\stackrel{\loon}{{\rm Re}\, \x \, \rightarrow + \infty}} \;
\frac{\l_0\p^2}{12N} \;+
\; {\cal O} (1) \;\; ,
\end{equation}
so that
\begin{equation}
{\rm Re} \; V(\p^2) \;
_{\stackrel{\loon}{{\rm Re} \, \x \rightarrow + \infty}} \;
\frac{\l_0\p^4}{24N} \;+
\; {\cal O} (\p^2) \;\; .
\end{equation}
Therefore the effective potential has a lower-bound for
$\l_0 > 0$
when ${\rm Re}\, \x \rightarrow + \infty$.  The
behavior exhibited by the cutoff theory in
(2.6)--(2.8) is in marked contrast to that of the renormalized vector
model, where \cite{002}
$$
{\rm Re} \; \x   \;\;\;
_{\stackrel{\frown\!\!\smile}{\p^2 \rightarrow \infty}} \;\;\;
\frac{-16\pi^2 (\p^2/N)}{\ln (\p^2/M^2)}
\vspace{-.2in}
$$
\hspace*{4.75in}{(renormalized theory)}\vspace{-.2in}
$$
{\rm Re} \;\; \frac{\pa V(\p^2)}{\pa\p^2} \;\;
_{\stackrel{\frown\!\!\smile}{\p^2 \rightarrow \infty}} \;\;\;
\frac{-8\pi^2 (\p^2/N)}{\ln (\p^2/M^2)}
$$
independent of the parameters of the theory, but with $M^2$ a
renormalization mass.  [These are eqs. (2.14) and (2.15) of Abbott, {\it
et al.} \cite{002}]   A comparison of (2.6)--(2.8) with the analogous
equations of the
renormalized theory (shown above) emphasizes the dramatic
difference between the behavior of the effective potential in the two
cases.  In the cutoff theory, the effective potential with $\l_0 > 0$
behaves classically as
${\rm Re}\, \x \rightarrow +\infty$, while in the renormalized theory,
the
effective potential is dominated by quantum effects in this limit.

{}From (2.5) we observe that the effective potential is everywhere complex
for $\x < - \L^2$, and that
\begin{equation}
{\rm Re} \:V \;
_{\stackrel{\frown\!\!\smile}{\x \rightarrow -\L^2_-}} \; -\infty \; .
\end{equation}
As a result of this behavior, we will eventually restrict the classical
external fields so that $\x > -\L^2$.   Before imposing such a
constraint, we first study (in the next section) the effective potential
without restrictions on $\x$ or the sign of $\l_0$.  The behavior of the
effective potential in this unrestricted case will be instructive, as it will
underline the limitations of the model.  We
will argue that it is plausible to require that $\l_0 > 0$ and $\x >
-\L^2$, and then will study the
consequences of the resulting restricted model.

Since the most important application of the model is to problems with
spontaneous symmetry breaking, we end this section by focusing on the
conditions for this to occur.  At a stationary point of the effective
potential, the right-hand side of (2.2) must vanish.  If the classical field
at this point satisfies $\p_a = <\P_a > \neq 0$, then \cite{001,002}
\begin{equation}
\x = 0
\end{equation}
for spontaneous symmetry breaking.  Combining (2.10) with
(2.4) implies that
\begin{equation}
\p^2 = <\P_a >^2 \; = \; - N \;
\left( \frac{6\m^2_0}{\l_0} \; + \; \frac{\L^2}{16\pi^2} \right)
> 0
\end{equation}
at the minimum of the potential.
Since in our three parameter model, $\m_0, \; \l_0$ and $\L$ are
finite, spontaneous symmetry breaking requires
\begin{equation}
\left( \frac{\m^2_0}{\l_0} \right) < -
\left( \frac{\L^2}{96\pi^2} \right) < 0 \;\; .
\end{equation}
For convenience define
the auxiliary quantity
\begin{equation}
\frac{\m^2}{\l} \; = \; \frac{\m^2_0}{\l_0} \; + \;
\frac{\L^2}{96\pi^2} \; .
\end{equation}
We call $(\m^2/\l )$ a ``dressed" parameter, in contrast with the
usual renormalized parameters of renormalized theories.  That is, since
$\L$
is finite, the bare quantities are dressed by interactions, even though
the bare quantities are themselves finite.  It is convenient to define
$$
v^2_0 = \frac{-6N\m^2_0}{\l_0}
\eqno{(2.14a)}
$$
and
$$
v^2 = \frac{-6N\m^2}{\l}
\eqno{(2.14b)}
$$
so that
\renewcommand{\theequation}{2.\arabic{equation}}
\setcounter{equation}{14}
\begin{equation}
< \P_a >^2 = v^2 \; = \; v^2_0 \; - \; \frac{N\L^2}{16\pi^2} > 0
\end{equation}
when the O(N) symmetry is  spontaneously broken.  [From (2.12) we
find the
usual condition, $\m^2_0/\l_0 < 0$ for spontaneous symmetry
breaking.]

\ni{\bf III. ~The Effective Potential}

In this section we discuss the effective potential of the cutoff vector
model first without placing any restrictions on $\x$ or $\l_0$.  Our
analysis will show that the unrestricted cutoff model is
inconsistent,
as the effective potential has no lower bound.  As a result, we will argue
that $\l_0 > 0$ and $\x > -
\L^2$ is required if the model is to be viable as an approximate
description of a strongly interacting scalar sector.

The effective potential $V(\p )$ can be evaluated from (2.5), using the
relation between $\x$ and $\p^2$ given by (2.4).  The analysis is
straightforward, and follows the strategy of Abbott, {\it et al.}
\cite{002}.  In so
doing, one must carefully examine (2.4) and (2.5) for multiple branches
of the potential.  Since the analysis is a bit tedious, and the details not
particularly illuminating, we present only the results in a series of
figures.

\ni{\bf A. ~~Unrestricted model}

Results for the case $\l_0 > 0$, together with $(\frac{6\m^2 N}{\l} )<0$
[for small and large values of the coupling $\l_0$ respectively] are
presented in
Figs. 1
and 2.  Fig. 3 gives $V(\p^2 )$ for $\l_0 >0$ and
$(\frac{6\m^2 N}{\l} ) > 0$.   In Figs. 4 and 5, $V(\p^2)$ is displayed
for $\l_0 < 0$ with  $(\frac{6\m^2 N}{\l} ) < 0$ and
$(\frac{6\m^2 N}{\l} ) > 0$ respectively.   Dotted lines in the figures
indicate Im $V \neq 0$.

We observe in Figs. 1 to 5 that in each case the effective potential is
multivalued, and has no lowest energy bound.  In each of the figures,
branch II, which corresponds to $\x < - \L^2$, is everywhere complex,
and lies below branch I for large $\p^2$.  Note that branch I
of our Figs. 4 and 5 are qualitatively similar to Figs. 2 and 3 of Abbott
{\it et
al.}   On the other hand branch I of our Figs. 1 to 3 behave qualitatively
like that of the classical theory.  In summary, we conclude from Figs.
1 to 5, that the cutoff vector model is not consistent if no restrictions
are placed on the model.
[See Sec. 7 for further comments on the consistency of cutoff-models.]

\ni{\bf B. ~~Restricted model}

We impose the restriction
\renewcommand{\theequation}{3.\arabic{equation}}
\setcounter{equation}{0}
\begin{equation}
\x > - \L^2
\end{equation}
on the external fields.  Clearly (3.1) eliminates branch II of $V(\p^2 )$
from consideration.  Branch II represents an instability which occurs
when the classical field $\x$ becomes too strong.  [It is useful to
remember that $\L^2$ is a Euclidean cutoff.]  We impose (3.1) and
show $V(\p^2 )$ in Figs. 6 and 7 for $\l_0 > 0$ and $6\m^2 N/\l < 0$
and $6\m^2 N/\l > 0$ respectively.  Figs. 8 and 9 give analogous
results
for $\l_0 < 0$.  Notice that now $V(\p )$ is single valued for $\l_0 >
0$, in contrast to the renormalized theory and is similar to that of the
classical theory, while for
$\l_0 < 0$ the effective potential qualitatively resembles the effective
potential of Abbott, {\it et al.}  [However, in the cutoff model
${\rm Re} \; V (\p^2) \;
_{\scriptstyle \stackrel{\loon}{\p^2 \rightarrow
\infty}}\; \frac{-|\l_0 | \p^4}{24N}$ when $\l_0 < 0$, while in the
renormalized
theory ${\rm Re} V(\p^2) \sim \frac{-4\pi^2\p^4}{\ln \p^2}$ for all
values of the coupling constant.]

In the next section we consider whether tachyons are present in the
restricted model.

\vspace{.2in}

\ni{\bf IV. ~Green's Functions and Tachyons}

In this section we consider the $\p -\x$ inverse propagator for the
cutoff vector model, and impose the restriction $\x > -\L^2$ on the
external classical fields.  Figures 6 to 9 describe the effective potential
for this case.  As noted in Sec. 3, when $\l_0 < 0$ the effective potential
has features which are  qualitatively similar to that of Abbott, {\it et
al.}, [{\it c.f.} their figs.
2 and 3].  Therefore, for reasons discussed in the Introduction and
ref. \cite{002}, we consider the cutoff model with $\l_0 < 0$ unsuitable
for
phenomenology.  Therefore we focus on the cutoff model with $\x > -
\L^2$ and $\l_0 > 0$ for the bulk of the paper.  We return to the
question of $\l_0 < 0$ when we consider the possibility of a
double-scaling limit in Sec. 6.

The $\Phi - \mbox{\boldmath $\x$}$ matrix inverse propagator in the
presence of external
fields $\p$ and $\x$ is [{\it c.f.} ref. \cite{002}, eqn. (4.2)].
\renewcommand{\theequation}{4.\arabic{equation}}
\setcounter{equation}{0}
\begin{equation}
D^{-1} (-k^2,\p , \x ) =
\left[ \begin{array}{cc}
(k^2 + \x )\d_{ab} & \p_a\\
\p_b & -3N\left( \frac{1}{\l_0} - \bar{B} (\x , k^2,\L^2 ) \right)
\end{array}
\right]
\end{equation}
for Euclidean momenta $k^2$.  The function
\begin{eqnarray}
\bar{B} (\x , k^2,\L^2 ) & = & - \frac{1}{6} \:
\int^\L \;  \frac{d^4p}{(2\pi )^4} \;
\frac{1}{(p^2 + \x ) [(k + p)^2 + \x ]} \nonumber \\
   & < & 0
\end{eqnarray}
is given by an integral over Euclidean momenta, and is real for $\x >
-  \L^2$.
The fact that the ``bubble" integral (4.2) is real for $\x > -\L^2$ is an
important aspect of the consistency of our approach, since this is the
same requirement for an acceptable effective potential.
The integration is made finite by a cutoff $\L$, with
some ambiguity in specifying a cutoff.  For simplicity we use a
sharp-cutoff in
$p$-space.  We have
verified by explicit calculation, that several different cutoff methods give
substantially
the same result.  In practice this is not a problem.  The essential point
is that \ul{all} reasonable cutoffs
insure that $\bar{B} < 0$ for Euclidean $k^2$.  This point was also
emphasized in ref. 17.  [By contrast, in the
renormalized theory, $\bar{B} (k^2)$ need not have a definite sign, as
it is made finite by a subtraction.]  On carrying out the
integration in (4.2) with a sharp cutoff, we find
\begin{eqnarray}
\bar{B} (\x , k^2,\L ) & = & - \; \frac{1}{96\pi^2}
\left\{ \ln \left( \frac{\L^2 + \x}{\x} \right) +  \right.\nonumber
\\[.2in]
&& + \; 2 \sqrt{\frac{k^2 +4(\L^2 + \x )}{k^2}} \;
\left[ 1 - \, \frac{2\L^2}{k^2 + 4 (\L^2 +\x )} \right] \ln
\left[ \frac{\sqrt{k^2} + \sqrt{k^2 + 4 (\L^2 + \x)}}
{2 \sqrt{\L^2 + \x}} \right] \nonumber \\[.2in]
&& \left.  - \; 2
\sqrt{\frac{k^2 +4\x}{k^2}} \, \ln \,
\left[ \frac{\sqrt{k^2} + \sqrt{k^2 + 4\x}}{2\sqrt{\x}} \right]
\right\}\;\; .
\end{eqnarray}
This expression differs from Abbott, {\it et al.}, since we have
assumed
neither $\x \ll \L^2$ nor $k^2 \ll \L^2$.  This will be important in
what follows.

\ni{\bf A. ~~Unbroken Symmetry}

The O(N) symmetry is unbroken if $< \P_a > = \p_a = 0$.  Since $\x
> 0$,  $\x
= m^2$ becomes the physical meson mass, and
\begin{equation}
D^{-1} (-k^2,0 , m^2 ) =
\left[ \begin{array}{cc}
(k^2 + m^2 )\d_{ab} & 0 \\
0 & -3N\left( \frac{1}{\l_0} - \bar{B} (m^2 , k^2,\L^2 ) \right)
\end{array}
\right]
\end{equation}
is the $\Phi - \mbox{\boldmath $\x$}$ inverse propagator for this case.

 When $\x > -\L^2,
\; \bar{B} (m^2, k^2, \L^2 ) < 0$ and real for $k^2$ Euclidean.
Therefore from (4.2)
\begin{equation}
\left[ \frac{1}{\l_0} - \bar{B} (m^2, k^2, \L^2 )\right] > 0
\end{equation}
for $\l_0 > 0$, independent of the details of the cutoff.  Hence, $D^{-1}$
does not vanish for Euclidean $k^2$, and tachyons are absent from the
model in this phase which is as expected from the effective potential,
Fig. 7.

\ni{\bf B. ~~Spontaneously Broken Symmetry}

The O(N) symmetry breaks spontaneously to O(N--1) if
\begin{eqnarray}
\langle \P_a \rangle^2 & = & \p^2 \; = \; v^2 \nonumber \\[.1in]
& = & - \; \frac{6N\m^2}{\l} \; > \; 0
\end{eqnarray}
and $\x = 0$.  The existence of tachyons depends on whether
$\det D^{-1} (-k^2, v, 0)$ vanishes for Euclidean $k^2$, {\it i.e.}, if
\begin{eqnarray}
\lefteqn{ \det D^{-1} (-k^2, v, 0) } \nonumber \\[.1in]
&& = k^2 \left\{ - 3N \left[ \frac{1}{\l_0}  - \bar{B}
(0, k^2, \L^2 ) \right] \right\} - v^2 \stackrel{\displaystyle ?}{=} 0
\end{eqnarray}
However, $-\bar{B} (0, k^2, \L^2 ) > 0$ for $k^2$ Euclidean,
independent of the details of the cutoff.  Therefore for
$\l_0 > 0$ and $v^2 > 0$, tachyons are absent from this phase, in
accord with the effective potential described by Fig. 6.  In fact
$$
\bar{B} (0, k^2, \L^2 )  =  -  \frac{1}{96\pi^2}
\left\{ \ln \left( \frac{\L^2}{k^2} \right)
 + 2 \sqrt{\frac{k^2 + 4 \L^2}{k^2}}
\left[ 1- \, \frac{2 \L^2}{(k^2 + 4\L^2 )} \right] \ln
 \left[ \frac{\sqrt{k^2} + \sqrt{k^2 + 4\L^2}}{2\L} \right]
\; \right\} \eqno{(4.8a)}
$$
$$
_{\stackrel{\loonl}{k^2 \gg \L^2}} \; - \;
\frac{1}{96\pi^2} \left\{  2 \left( \frac{\L^2}{k^2} \right) +
{\cal O} (\L^4/k^4) \right\} < 0
\eqno{(4.8b)}
$$

$$
_{\stackrel{\loonl}{k^2 \ll \L^2}} \; - \;
\frac{1}{96\pi^2} \left\{ \ln \left( \frac{\L^2}{k^2} \right) + 1 +
{\cal O} ( k^2/\L^2 ) \right\} < 0
\eqno{(4.8c)}
$$
[Notice that (4.8c) changes sign, if it is naively used for $k^2 \gg
\L^2$.
This unjustifiable use of (4.8c) for large $k^2$ violates the general
requirement that $\bar{B} < 0$.  If (4.8c) were to be used for all $k^2$,
one
would conclude erroneously that there was a
tachyon in this phase.]

We conclude that the cutoff vector model with $\l_0 > 0$ and $\x > -
\L^2$ is free of tachyons to leading order in 1/N.  It would be
interesting to see if this feature persists in higher orders of 1/N.

\newpage

\ni{\bf C.} ~$\mbox{\boldmath $\l_0 < 0$}$

We briefly summarize the issue of tachyons for $\l_0 < 0$.  The
effective
potential is given by Figs. 8 and 9 for our model restricted to
$\chi > -\l^2$.  Notice the similarity of our Figs. 8 and 9 to the
effective potentials found by Abbott, {\it et al.} \cite{002} in their Figs.
2 and 3.  When $\l_0 < 0$ and $(\m^2/\l ) < 0$, (4.7) is the appropriate
equation
for this issue.  If $\langle \Phi \rangle^2 > 0$, what is relevant is the
upper-branch
of Fig. 8.  It is straightforward to show that tachyons are \ul{always}
present for vacuua chosen on the upper branch of Fig. 8.  Similarly one
can ask whether tachyons are present when $\l_0 < 0$ and $(\m^2/\l
) > 0$.
[Figure 9 is now appropriate.]  Tachyons will be present if
(for $\l_0 < 0$)
\renewcommand{\theequation}{4.\arabic{equation}}
\setcounter{equation}{8}
\begin{equation}
-\bar{B} (m^2 , \; k^2 , \; \L^2) \: \stackrel{?}{=} \:
\left| \; \frac{1}{\l_0} \; \right| \; .
\end{equation}
Since
\begin{equation}
-\bar{B} (m^2, \; 0, \; \L^2) > - \bar{B}
(m^2, \; k^2, \; \L^2) > 0
\end{equation}
for $k^2$ Euclidean, tachyons will be absent if
\begin{equation}
0 < -\bar{B} (m^2, \; 0, \; \L^2) <
\left| \; \frac{1}{\l_0} \; \right| \; ,
\end{equation}
{\it i.e.}, if
\begin{equation}
\frac{1}{96\pi^2}
\left[ \ln \left( \frac{\L^2 + \x}{\x} \right) -
\left( \frac{\L^2}{\L^2 + \x} \right) \right] <
\left| \; \frac{1}{\l_0} \; \right| \; .
\end{equation}
Again, a straightforward analysis analogous to that of Abbott, {\it et
al.}, shows that the \ul{upper-} \ul{branch} of Fig. 9 \ul{always} leads
to tachyons.

What about the lower-branch of Figs. 8 and 9?  In that case the same
analysis shows that the tachyons are \ul{always} absent.  In fact the
branch-point appears at $\phi^2 = 0$ in Figs. 8 or 9 just  when there
is
a zero-mass O(N) singlet bound state.  We return to the case $\l_0 < 0$
in Sec. 6 when we discuss a possible double-scaling limit.

\vspace{.2in}

\ni{\bf V. ~Physical Properties}

\ni{\bf A. ~~The Spectrum}

We have seen that when $\l_0 > 0$ and $<\P_a >^2 = v^2 = -6N\m^2
/\l$, the theory has a spontaneously broken symmetry, without
tachyons being present.   This phase has (N--1)  massless
Goldstone bosons transforming as an O(N--1) vector.  One can explore
the spectrum in the O(N--1) singlet
sector by continuing det $D^{-1}$ to Minkowski momenta.
An O(N--1)
singlet resonance,
will occur if
\renewcommand{\theequation}{5.\arabic{equation}}
\setcounter{equation}{0}
\begin{equation}
\left[ \frac{1}{\l_0} - Re \; \bar{B}
(0, -s_r, \L^2 ) \right] =
\frac{1}{3N} \;
\left( \frac{v^2}{s_r} \right)
\end{equation}
has a solution.  [If the width of the resonance can be neglected,
$\sqrt{s_r}$ will be the mass of the resonance.]  Since
$$
- {\rm Re} \; \bar{B} (0,-s,\L^2 ) \; \:
_{\stackrel{\loon}{s\ll \L^2}} \;\:
\frac{1}{96\pi^2} \;
\left\{ \ln \left( \frac{\L^2}{s} \right)  +
\ldots \right\} > 0 \eqno{(5.2a)}
$$
and
$$
- {\rm Re} \; \bar{B} (0,-s,\L^2 ) \;\:
_{\stackrel{\loon}{s \rightarrow 4 \L^2_-}} \;\:
\frac{-\L}{96\pi\sqrt{4\L^2-s}} \;
+ \ldots \;\;\;  < 0 \eqno{(5.2b)}
$$
equation (5.1) will always have a solution.

First consider the possibility that $s_r \ll \L^2$.  Then
the width is negligible, and
we estimate
\renewcommand{\theequation}{5.\arabic{equation}}
\setcounter{equation}{2}
\begin{equation}
s_r \simeq \frac{\l_0 (v^2/3N)}
{\left[ 1+\frac{\l_0}{96\pi^2} \ln (\L^2/s_r) \right]} \ll \L^2
\end{equation}
which is consistent if $\frac{\l_0}{3N} \left( \frac{v^2}{\L^2} \right) \ll
1$.  If further
$\frac{\l_0}{96\pi^2} \, \ln \left( \frac{\L^2}{s_r} \right) \ll 1$, then
\begin{equation}
s_r \simeq \frac{\l_0 \, v^2}{3N}
\end{equation}
which agrees with the weak-coupling, semi-classical evaluation of the
scalar
mass from (2.1), \ul{except} that $v^2_0$ has now been dressed to
$v^2$.

We do not have an analytic estimate of $s_r$ in the general case, so
that detailed results depend on numerical evaluation.  Notice from (4.8)
that $\bar{B} (0,-s,\L^2)$ only depends on $(s/\L^2)$.  Therefore, the
bound-state equation
depends on three dimensionless parameters $( s_r/\L^2 )$,
$(v^2/N\L^2)$, and $\l_0$, with (5.1) providing one relation between
them.
According to (2.1) the large N limit is taken by keeping $\l_0$ and $v^2
/N$ fixed.
Physically $v^2/N$ characterizes the broken symmetry
scale, while $\L^2$ characterizes the scale of ``new physics."
With these fixed, one can explore $(s_r/\L^2
)$ vs. $\l_0$, as $\l_0$ varies from weak to strong coupling.

We plot $-96\pi^2 \:{\rm Re}\: \bar{B} (0, -s, \L^2 )$ versus
$x = s/\L^2$  in Fig. 10 as a solid line.  In the same figure we also
plot $ 96\pi^2 [ \frac{1}{3N} (\frac{v^2}{\L^2}) \frac{1}{x} -
\frac{1}{\l_0}]$
for a ``typical" case, shown as a dotted line.  Changes in $\l_0$ merely
raise or lower the dotted curve.  Figure 10 enables us to understand the
qualitative behavior of solutions to (5.1).  It is clear from the figure that
there are always \ul{two} solutions of (5.1); one with
$s_r/\L^2 < 1$, and the other with $\bar{s}/\L^2 > 1$.  The solution
$\bar{s}$ should \ul{not} be considered a prediction of the model, as
$\bar{s}$ \ul{always} appears above the cutoff energy of the model.
[We comment on the interpretation of this unphysical resonance at the
end of this section.]  With this understanding, we conclude that the
model predicts one resonance with $s_r < \L^2$ in the O(N--1) singlet
channel, which is the Higgs meson in typical
applications.  In the limit where $\frac{\l_0}{3N} (\frac{v^2}{\L^2}) \ll
1$, the Higgs mass is given by (5.3) [or (5.4) if appropriate].  Figure 10
makes it clear that as $\l_0$ is increased, with $v^2/N$ and $\L$
fixed, the Higgs mass and width increase, but with $s_r < \L^2$
always.  However, as $\l_0$ increases, the width increases as
$(s_r)^{3/2}$, so is eventually no longer negligible.  We consider this
situation in the next subsection.

\ni{\bf B. ~~Scattering Amplitudes}

The $\Phi - \mbox{\boldmath $\x$}$ matrix inverse propagator is
easily inverted.  In the
Minkowski region
\begin{equation}
D_{\x\x} (0,s^2,\L^2 ) =
\frac{(s/3N)}{s\left[ -\frac{1}{\l_0} + \bar{B} (0,-s,\L^2)\right]
+ v^2/3N}
\end{equation}
\begin{equation}
_{\stackrel{\loonl}{s \ll \L^2}} \;
\frac{(s/3N)}{s
\left\{ - \frac{1}{\l_0} - \frac{1}{96\pi^2} \left[ \ln
\left( \frac{\L^2}{s} \right) + i\pi \right] \right\} + v^2/3N}
\end{equation}
A typical example is the meson-meson scattering amplitude in the
O(N--1)
singlet sector is
\begin{equation}
A(s,t,u) = D_{\x\x} (s) + D_{\x\x} (t) + D_{\x\x} (u)
\end{equation}
which has an $s$-channel resonance, with a width proportional to
$(s_r)^{3/2}$, and a scalar particle exchange in the $t$ and
$u$-channels.  In
physical applications this would correspond to longitudinal $Z_0 -Z_0$
scattering \cite{008}, for example.
The amplitudes for the scattering of longitudinal $W$'s for other
channels are given by eq. (1) of Naculich and Yuan \cite{007}.  If
$s/\L^2$ is not $\ll 1$, then one must use (5.5) together with (4.8$a$)
continued to Minkowski momenta, rather than the approximate
expression (5.6).  Then the resonance mass is computed from the
position of the peak in Im $D_{\x\x}(s)$.  A table of resonance masses
vs. coupling constant, for $\L = 1$ Tev and $\L = 4$ Tev is given in
Table I.  The
weak-coupling approximation, with the width ignored in (5.1) is
satisfactory for $(\l_0/16\pi^2) \leq 1.3$ for $\L = 1$ Tev and for
$(\l_0 /16\pi^2) \leq 2.5$ for $\L = 4$ Tev.  The width $\G \sim 300$
Gev for $\L = 1$ Tev and $\G = 240$ Gev for $\L = 4$ Tev when
$(\l_0/16\pi^2) = 0.95$, and varies as $(s_r)^{3/2}$.

{}From (4.8$a$). (5.5) and Fig. 10 we see that cross-sections go through
a second unphysical resonance.  [See Sec. 5a above.]  We identify this
with the \ul{triviality scale} $\sqrt{\bar{s}}$ discussed in the
literature \cite{005,006,007}.  In the weak coupling limit, one finds from
(5.2$b$) that $(4\L^2 - \bar{s})/\L^2 \simeq (\frac{\l_0}{96\pi})^2$.
As $\l_0$ increases, $\bar{s}$ decreases towards $\L^2$ from above,
while the Higgs (mass)$^2$ $s_r$ increases towards $\L^2$ from
below.

\vspace{.1in}

\noindent{\bf VI. ~Is There a Double-Scaling Limit?}

There has been considerable interest in the double scaling limit for
matrix models \cite{009}, for which one considers the correlated limit
$N \rightarrow \infty$ and $g \rightarrow g_c$, where $g_c$ is
a critical value of a coupling constant.  Unfortunately this approach to
quantum gravity meets a c=1 barrier.  It has been suggested that one
consider a double-scaling limit for O(N) models for dimensions
D$\geq$2 \cite{010}, as the Feynman diagrams of such theories do not
fill a surface, but rather have a branched structure.  It was hoped that
such considerations would give some indications on how the c=1 barrier
might be surmounted.  However, it has been shown \cite{011} that at
the critical point of the D=4 renormalized vector model, the effective
potential is everywhere complex, which implies that there is no
double-scaling limit for this model.  Similar results have been found for
D=2
and D=3 \cite{012,013}.

It is interesting to reexamine the possibility of a double scaling limit for
the cutoff vector model as an application of our study of the effective
potential.   We require a zero-mass bound-state in the O(N) singlet
channel.  An analysis similar to that of Sec. 4A shows that this is not
possible for $\l_0 > 0$, and therefore we turn to $\l_0 < 0$ as in Sec.
4C.  A zero-mass O(N) singlet bound-state implies that

\newpage

\renewcommand{\theequation}{6.\arabic{equation}}
\setcounter{equation}{0}
\begin{eqnarray}
\frac{1}{\l_0} & = & \bar{B} (\x , \; 0, \; \L^2) \\ \nonumber
& = & -\frac{1}{96\pi^2}
\left\{ \ln \left( \frac{\L^2 + \x}{\x} \right)  -
\left( \frac{\L^2}{\L^2 + \x} \right) \right\} < 0
\end{eqnarray}
Equation (6.1) can be combined with the gap equation (2.4) to eliminate
the logarithm term, with the result
\begin{equation}
\left( \frac{\x}{\L^2 + \x} \right) = \left( \frac{96\pi^2}{\L^2} \right)
\left( \frac{\m^2}{\l} \right) > 0 \; .
\end{equation}
Note that now $\x$ is single-valued just when the theory has a
zero-mass O(N)  singlet bound-state.  The solution to the gap equation
is now unique.  This occurs precisely where the
upper and lower branch of Fig. 9 meet.  As a consequence, the effective
potential $V(\phi )$ is \ul{everywhere complex} for this choice of
parameters.  Therefore, there is \ul{no} double-scaling limit for the
cutoff model for precisely the same reason \cite{011} that a
double-scaling limit
is absent in the renormalized theory.  Our detailed analysis of the
effective potential agrees with a general argument of Moshe \cite{013}.
The discussion of this section reiterates that the upper-branch of Figs.
8 or 9 always
leads to amplitudes with tachyons, while the lower-branch does not, as
the zero-mass bound-state divides the two branches.

\vspace{.1in}

\ni{\bf VII. ~Concluding Remarks}

In this paper we have studied a cutoff version of the O(N) $\l\phi^4$
vector
model to leading order in the 1/N expansion.   An important aspect of
our work is
a detailed analysis of the effective potential of the model; a topic not
studied in previous work.  We found that the
effective potential was surprisingly rich in structure.  Most notably
there is a \ul{disjoint} branch of the effective potential which is
everywhere complex, coming from $\x < - \L^2$ for all values of the
parameters of the model, and which has no lowest energy [see Figs. 1
to 5].

It is known for a very long time \cite{014} that a cutoff-model should
have some \ul{fundamental} difficulties, as higher-derivative or
non-local cutoff theories suffer either from ghosts and absence of
unitarity ({\it e.g.}, Pauli--Villars theories), or no lowest energy bound.
In fact one can convert an indefinite metric theory with ghosts to one
with positive metric and no lowest energy by a change in the definition
of the adjoint operator, {\it e.g.}, $A^+ = (-1)^n \: A^*$, where $n$ is
the ghost number operator, and $A^+ (A^*)$ is the new (usual) adjoint
operator.  Attempts at making indefinite theories completely consistent
\cite{015} have failed \cite{016}.  Therefore, \ul{any} cutoff O(N) vector
model should have \ul{some} fundamental difficulty.  The models
studied by Heller, {\it et al.}, \cite{017} have ghosts, but do have a
lowest energy bound, while our model does \ul{not} have ghosts, but
fails to have a lowest energy state.  The analysis of Pais and Uhlenbeck
\cite{014}, and the above remarks strongly suggest that these two
difficulties are in fact two different aspects of the \ul{same} problem,
which may be related by a change in the adjoint operator, as mentioned
above.

An essential result of our cutoff model is that the branch of the
effective potential which has no lowest energy and is everywhere
complex is \ul{disjoint} from the rest of the potential, at least to leading
order in 1/N.  It is this feature which allowed us to restrict our
attention to those branches of the effective potential for which $\x > -
\L^2$.

Thus analysis of the effective
potential showed that one must restrict the composite classical field
$\x$ to satisfy $\x > -\L^2$, where $\L$ is the fixed cutoff
mass-scale of the model, and the bare coupling constant $\l_0 > 0$, in
order for the model to be consistent to leading order in 1/N.  These
same constraints guarantee
that the model has a phase with spontaneously broken symmetry, free
of tachyons.  In this phase the O(N--1) singlet channel of meson-meson
scattering has two resonances, one physical with (mass)$^2 = s_r <
\L^2$, and the other unphysical since its (mass)$^2 = \bar{s} > \L^2$.
In applications the physical state is usually identified with the Higgs
boson, while we identify the mass of the unphysical state
$\sqrt{\bar{s}}$ with the triviality scale, since for $s \geq \bar{s}$ the
model is certainly physically unacceptable.  That is, features of the
model which appear for $s > \L^2$ are not to be considered
predictions of the model, but merely artifacts of the cutoff model.

For weak coupling defined by $\frac{\l_0}{3N} (\frac{v^2}{\L^2}) \ll
1$, the Higgs boson (mass)$^2$ satisfies $s_r \ll \L^2$, {\it c.f.} (5.3).
If further, logarithmic corrections are small, then $s_r$ agrees with the
semi-classical estimate, but with dressed vacuum expectation value, {\it
c.f.} (5.4).  In weak-coupling, the triviality mass satisfies
$\bar{s} \simeq 4\L^2$.  As the coupling $\l_0 > 0$ increases, the
Higgs mass and width increases, with $s_r < \L^2$.  At the
same time, the triviality mass decreases toward $\L$, with
$\bar{s} > \L^2$.  Thus, we have
$0 < s_r < \L^2 < \bar{s} < 4\L^2$.

The behavior of the Higgs mass and width, and triviality
mass-scale are in qualitative agreement with previous studies
\cite{005,006,017}.  Certain
features of our particular version of the model should be emphasized
however.  ~1)
Study of the effective potential leads to the constraints $\x
> -\L^2$ and $\l_0 > 0$.  ~2) The constrained model is tachyon free,
and consistent for  $s < \L^2$.   Thus, we were able to use
very general properties of the
bubble integral, whereby $\bar{B} (\x ,k^2,\L^2 ) < 0$ and real for
$\x > -\L^2$ [{\it c.f.} (4.2)],  to demonstrate the absence of tachyons
in the restricted model.  This conclusion therefore does not depend on
the details of the cutoff procedure.

There are definite advantages in having a tachyon free formulation of
the cutoff vector model.  For example, one can hope to go beyond the
leading order in 1/N, so as to calculate 1/N$^2$ corrections.  If tachyons
are present in the 1/N approximation as isolated unphysical states, they
will then appear in loop-corrections in the next order in 1/N.  As a
consequence, the effective potential will be everywhere complex in
non-leading orders of 1/N [{\it c.f.} ref. \cite{002} and Root, ref.
\cite{001}],  rendering
the model inconsistent.   If tachyons are absent in leading order, then
there is the possibility that the theory will remain consistent in
higher
orders of 1/N.   Therefore, a tachyon free formulation of the cutoff O(N)
model to leading order in 1/N is very attractive from a theoretical point
of view.

There are still some outstanding questions which deserve further
scrutiny.  It is not known whether the tachyon free ground-states of the
$\l_0 > 0$ models are metastable or not.  To resolve this issue, a
calculation at least to the next order in 1/N will be necessary.  It is
promising that the ground-state which is tachyon free and the branch
with $\x < -\L^2$ (which has no lowest energy) are \ul{disjoint} in the
effective potential.  The issue therefore is whether or not the
tachyon free ground-state will decay to the lower branch in a finite
order of the 1/N expansion.  In this context, recall that the tachyon free
vacuum found by Abbott, {\it et al.}, \cite{002} \ul{was} metastable in
the renormalized theory.

It might be thought that this issue might be avoided by choosing a
model with a Pauli--Villars cutoff (say) which does have a lowest
energy, but has ghosts.  Since the ghosts represent a failure of
unitarity, it is not clear how this will be manifest in
higher orders in 1/N.  As we have already remarked, a model with
ghosts or a model with no lowest energy seem to be two different
aspects
of the same problem of cutoff theories \cite{014}.  Therefore, we expect
that the issue of a metastable ground-state applies to both
versions of the model.  It cannot be avoided.

\vspace{.2in}

\ni{\bf Acknowledgement}

We wish to thank Professor Steve Naculich for discussions and reading
the manuscript.

\vspace{.2in}

\noindent{\bf NOTE ADDED:}

After this paper was initially submitted for publication we became
aware of the work by Heller, Neuberger, and Vranas \cite{017}.  These
authors study related cutoff O(N) models in the large N limit.  We wish
to thank Professor Neuberger for bringing this to our attention.   Their
model is not identical to the $\l\p^4$ theory considered here, as they
add dimension 6 and 8 operators to the theory and modify the kinetic
term.  They consider a class of Pauli--Villars regularizations and lattice
regularizations as well.  The phenomenological aspects of our model are
consistent with the conclusions of Heller, {\it et al.} \cite{005,017}.
It should be emphasized that these authors do not study the effective
potential or the double-scaling limit, which were presented in this
paper.  However, it is likely that their analysis of the Higgs mass and
triviality mass-scale is numerically more accurate for applications to
phenomenology.

Some authors have expressed the opinion that further study of the large
N model would be of ``progressively diminishing interest."  We do not
share this view.  Our work shows that a better understanding of the
model is made possible by investigation of the effective potential, which
in turn suggests issues for further study.  In fact we believe that even
reformulation of known results can lead to new insights.

\newpage

\renewcommand{\baselinestretch}{2}
\small
\normalsize

\vspace*{.5in}

\begin{center}
\begin{tabular}{|c|c|c|}
\hline
& \multicolumn{2}{c|}{$\sqrt{s_r}$ in Gev} \\
\hline
{}~~~~~$\l_0/16\pi^2$~~~~~ & for $\L = 1$ Tev & ~~~~~for $\L = 4$
Tev~~~~~ \\
\hline
0.02           &   100            &  80 \\
\hline
0.2            &   300            &  280 \\
\hline
0.4            &   410            &  390 \\
\hline
0.6            &   510            &  460 \\
\hline
0.95           &   610            &  520 \\
\hline
1.30           &   700            &  560 \\
\hline
1.9            &   860            &  640 \\
\hline
2.5            & resonance        &  660 \\ [-.2in]
\hrulefill     &                  & \hrulefill\\[-.18in]
3.2            & not well separated &  680 \\[-.2in]
\hrulefill     &                  & \hrulefill \\[-.18in]
3.8            & ~~~from triviality resonance~~~ &  680 \\[.1in]
\hline
\end{tabular}
\end{center}

\renewcommand{\baselinestretch}{1}
\small
\normalsize

\begin{center}
{\bf Table I}
\end{center}

\begin{quotation}\noindent\underline{Table Caption}:  Approximate
resonance mass
$\sqrt{s_r}$ in Gev vs. coupling constant $\l_0/16\pi^2$ for cutoffs $\L
= 1$ Tev and $\L = 4$ Tev.
\end{quotation}

\newpage

\ni{\bf Figure Captions}

\ni{\bf Fig. 1:} ~Effective potential for $\bar{\l}_0 > \l_0 > 0$,
$\frac{6\m^2N}{\l} < 0$, and no restriction on the background field
$\x$.  The parameter $\bar{\l}_0$ specifies an upper-limit for $\l_0$
for which branch II remains below branch I for all $\p^2$ beyond the
broken symmetry vacuum of branch I.
Branch I (II) comes from $\x > -\L^2 (\x < -\L^2)$.  This case
describes broken symmetry and weak coupling.  Dotted lines indicate
regions where the potential is complex.

\vspace{.1in}

\ni{\bf Fig. 2:} ~Same as Fig. 1, except for strong-coupling
$\l_0 > \bar{\l}_0  > 0$

\vspace{.1in}

\ni{\bf Fig. 3:} ~Effective potential for $\l_0 > 0$, $\frac{6\m^2N}{\l}
> 0$, and no restriction on the background field $\x$.

\vspace{.1in}

\ni{\bf Fig. 4:} ~Effective potential for $\l_0 < 0$, $\frac{6\m^2N}{\l}
< 0$, and no restriction on the background field $\x$.  Branch I (II)
comes from $\x > -\L^2 (\x < -\L^2 )$, with each split into 2
subbranches.

\vspace{.1in}
\ni{\bf Fig. 5:} ~Same as Fig. 4, except
$\frac{6\m^2N}{\l} > 0$.
\vspace{.1in}

\ni{\bf Fig. 6:} ~Effective potential for $\l_0 > 0$,
$\frac{6\m^2N}{\l} < 0$, with the restriction $\x > -\L^2$.

\vspace{.1in}

\ni{\bf Fig. 7:} ~Same as Fig. 6, but with $\frac{6\m^2N}{\l} > 0$.

\vspace{.1in}

\ni{\bf Fig. 8:} ~Same as Fig. 6, except $\l_0 < 0$.

\vspace{.1in}

\ni{\bf Fig. 9:} ~Same as Fig. 7, except $\l_0 < 0$.

\vspace{.1in}

\ni{\bf Fig. 10:} Graph of $-(96\pi^2 ) {\rm Re} \bar{B}(0, -s, \L^2)$
versus $x = s/\L^2$, shown as the solid line, the quantity $96\pi^2
[\frac{1}{3N} (\frac{v^2}{\L^2}) \frac{1}{x} - \frac{1}{\l_0} ]$ is also
plotted as a dotted line for a typical set of values of the parameters.
Changes in $\l_0$ shifts the dotted curve up or down.

\end{document}